\documentclass[english,conference]{IEEEtran}
\usepackage[T1]{fontenc}
\usepackage[latin9]{inputenc}
\usepackage{babel}
\usepackage{array}
\usepackage{amsmath}
\usepackage{amssymb}
\usepackage{stmaryrd}
\usepackage{graphicx}
\usepackage[unicode=true,
 bookmarks=false,
 breaklinks=false,pdfborder={0 0 1},backref=false,colorlinks=false]
 {hyperref}
\hypersetup{pdftitle={\@shorttitle},
 pdfauthor={\@shortauthor},
 pdfsubject={\@Subject},
 pdfkeywords={Computer Graphics Forum, EUROGRAPHICS},
   colorlinks,linkcolor=blue,citecolor=blue,urlcolor=blue,    bookmarks=false,    pdfpagemode=UseNone}

\makeatletter

\providecommand{\tabularnewline}{\\}

\IEEEoverridecommandlockouts
\usepackage{amsmath,amssymb,amsfonts}
\usepackage{algorithmic}
\usepackage{graphicx}
\usepackage{textcomp}
\usepackage{xcolor}
\def\BibTeX{{\rm B\kern-.05em{\sc i\kern-.025em b}\kern-.08em
    T\kern-.1667em\lower.7ex\hbox{E}\kern-.125emX}}
\usepackage[noadjust]{cite}

\usepackage{tikz,tkz-tab,tkz-graph}

\usepackage{balance}% for  \balance command ON LAST PAGE  (only there!)

\usepackage{tikz,tkz-tab,tkz-graph}
\usepackage{relsize}
\usepackage{amsthm}

\theoremstyle{plain}

\usetikzlibrary{arrows,automata}
\usetikzlibrary{positioning, fadings,snakes,calc,fadings,shapes.arrows,shadows,backgrounds}
\usetikzlibrary{shapes.geometric}
\tikzset{
    state/.style={
           rectangle,
  fill=#1!5!white,
           draw=#1, very thick,
           minimum height=2em,
           inner sep=2pt,
           text centered,
           },
    highlight/.style={
           rectangle,
  fill=#1!50!white,
           rounded corners,
           draw=#1, very thick,
           minimum height=2em,
           inner sep=2pt,
           text centered,
           },
coeff/.style={
           circle,
           draw=black, very thick,
           minimum height=2em,
           inner sep=2pt,
           text centered,
           },	
ptNode/.style={circle, fill=black,thick, inner sep=2pt, minimum size=0.2cm}	,
square/.style={regular polygon,regular polygon sides=4},
ptNodeSq/.style={square, fill=black,thick, inner sep=2pt, minimum size=0.2cm},	
}

\tikzset{
    invisible/.style={opacity=0},
    visible on/.style={alt=#1{}{invisible}},
    alt/.code args={<#1>#2#3}{%
      \alt<#1>{\pgfkeysalso{#2}}{\pgfkeysalso{#3}} % \pgfkeysalso doesn't change the path
    },
  }

\tikzset{fontscale/.style = {font=\relsize{#1}}
    }

\tikzset{
    *|/.style={
        to path={
            (perpendicular cs: horizontal line through={(\tikztostart)},
                                 vertical line through={(\tikztotarget)})
            -- (\tikztotarget) \tikztonodes
        }
    }
}

\usepackage{colortbl}
\definecolor{lightgray}{gray}{0.9}

\definecolor{S_purple}{RGB}{204, 0, 204}
\definecolor{S_brique}{RGB}{204, 51, 0}
\definecolor{S_petrol}{RGB}{0, 102, 153}
\definecolor{S_green}{RGB}{0, 153, 0}

\makeatother

\begin{document}
\title{The HyperBagGraph DataEdron: An Enriched Browsing Experience of Multimedia Datasets}
\author{
	\IEEEauthorblockN{Xavier Ouvrard} 	\IEEEauthorblockA{\textit{University of Geneva / CERN} \\ Meyrin, Switzerland \\ xavier.ouvrard@cern.ch}
\and
\IEEEauthorblockN{Jean-Marie Le Goff} 	\IEEEauthorblockA{\textit{CERN} \\ Meyrin, Switzerland \\ jean-marie.le.goff@cern.ch} 
	\and 
	\IEEEauthorblockN{Stéphane Marchand-Maillet} 
	\IEEEauthorblockA{\textit{CUI}   \textit{University of Geneva}\\ Carouge, Switzerland \\ stephane.marchand-maillet@unige.ch}  
}
\IEEEoverridecommandlockouts     
%\IEEEpubid{978-1-5386-7021-7/18/\$31.00~\copyright2018~\textrm{IEEE}}
\maketitle
\begin{abstract}
Traditional verbatim browsers give back information in a linear way according to a ranking performed by a search engine that may not be optimal for the surfer. The latter may need to assess the pertinence of the information retrieved, particularly when s$\cdot$he wants to explore other facets of a multi-facetted information space. For instance, in a multimedia dataset different facets such as keywords, authors, publication category, organisations and figures can be of interest. The facet simultaneous visualisation can help to gain insights on the information retrieved and call for  further searches. Facets are co-occurence networks, modeled by HyperBag-Graphs - families of multisets - and are in fact linked not only to the publication itself, but to any chosen reference. These references allow to navigate inside the dataset and  perform visual queries. We explore here the case of scientific publications based on Arxiv searches.
\end{abstract}
\begin{IEEEkeywords} hyperbag-graph, scientific publication searching, knowledge discovery, visual queries \end{IEEEkeywords}

\section{Introduction}

Revealing important information contained in Big Data calls for powerful
analytical and visualising tools to reveal important information it
contains. In an information space, meaningfull information can be
regrouped by hierarchical classification or - non exclusive - by semantically
cohesive categories that are combined to express concepts \cite{ranganathan1962elements}.

When browsing a textual database, traditional verbatim browsers give
back linear information in the form of a ranked list with a short
description of the references. To increase the pertinence of this
information, the surfer has often to perform new search either by
refining the original keywords s$\cdot$he used or by using other
pertinent queries that can help her$\cdot$him to refine the retrieved
information.

In fact, in most cases, the information space is multifacetted. Those
facets are not independent but are linked by the physical entities
contained in the search output. Choosing a type of reference common
to these entities enables the construction of a network of co-occurences
for each facet. Navigating to reveal information between the different
facets helps. For instance, in the case of scientific publications
different information of interest for the scientist can be regrouped
around the publication reference: the authors, the organization(s)
they belong, the countries of these organisations, the main keywords
of an article or the figures inside the publication. All this metadata
can give insights into the information space and as most of the articles
in the database can be systematically regrouped to get meaningfull
information. Choosing as reference the article itself, the facets
can show different co-occurrence networks such as co-authors, co-keywords
and co-categories. The interaction between the different facets is
ensured by the links between them. The choice of the reference type
is not unique and can be any convenient metadata type in the dataset.
Co-occurences can contain repetitions or an individual weighting:
multisets, and particularly natural multisets, allow them while sets
don't.

We propose in this article another way to explore an information space
by using hyper-bag-graphs (hb-graphs for short), an extension of hypergraphs
to families of multisets. In \cite{ouvrard2019exthbgraphdiffusion},
we show that hb-graphs enhance an exchange-based diffusion over co-occurence
networks, providing a fine vertex and hb-edge ranking, accounting
for a hb-edge based weighting of vertices.

This paper presents a hyper-bag framework of co-occurence networks
extending the visualisation part of the hypergraph framework sketched
in \cite{Ouvrard2018ENDMHypergraphMod} to hb-graph . This framework
supports browsing of an information space and dataset visual queries.
The framework is validated from a theoretical point of view and with
a use case. We have implemented a 2.5D interface to visualise different
facets of the Arxiv information space and perform visual queries.

Section II lists the related work and the mathematical background.
Section III presents the hb-graph framework. Section IV gives results
and Section V concludes and addresses future work.\IEEEpubidadjcol

\section{Related work and mathematical background}

\label{sec:Background}

\subsection{Information space discovery}

Discovering knowledge in an information space requires to gather meaningful
information, either hierarchically or semantically organized. Semantic
provides support to the definition of facets within an information
space \cite{ranganathan1962elements}.

Navigation and visualisation of information spaces have been achieved
by several authors in many different ways. In \cite{dork12PivotPaths:StrollingthroughFacetedInformationSpaces},
a pivot is used to stroll between three facets of the information
space; the approach is limited to the visualisation of a small amount
of pivots at the same time. This visualisation is based on a tri-partite
graph. In \cite{zhao2013interactive}, an interactive exploration
of implicit and explicit relations in faceted datasets is proposed.
The space of visualisation is shared between different metadata with
cross findings between metadata, partitioning the space in categories.

\cite{agocs2017interactive} proposes a visual analytics graph-based
framework that reveals this information and facilitates knowledge
discovery. It provides insights into different facets of an information
space based on user-selected perspectives. The dataset is stored as
a labelled graph in a graph database. Choosing a perspective as reference
and a facet as dimension, paths of the labelled graph are retrieved
with same dimension extremities going through reference vertices.
Visualisation comes in the form of navigable node-link graphs: edges
materialise common references between vertices and are seen as pairwise
collaboration between two vertices.

\subsection{Co-occurence networks}

\label{sec:Related works}

Data mining is only one step in the knowledge discovery processing
chain \cite{han2011data}. If numerical data allows rich statistics
on the instances, non numerical data mining consists often in summarizing
data as occurrences. Some other approaches exist. Often they consist
in regrouping data instances through similarities using techniques
such as $k$-nearest neighbours \cite{friedman1997bias}: using a
threshold, links are established between different occurrences. These
approaches relies on similarities: the curse of dimensionality is
nonetheless a limiting factor in their use \cite{weber1998quantitative},
even if some techniques exist to limit its impact \cite{indyk1998approximate,aggarwal2001surprising}.
A last way of finding occurences is to retrieve links through the
dataset itself.

If the dataset reflects existing links the job is easier since an
inherent network can be built through the data instances. A typical
example is with group of friends in social networks. When the links
exist, the collaborations are derived from them. Nonetheless, links
are often neither direct nor tangible: in this case occurences need
to be built or process from the dataset.

A dataset can be a set of physical references, stored as rows in traditional
relational databases. Each physical reference has a metadata instance
attached to it. Some of the types of the metadata instances are of
interest for visualisation and some for processing additional information.
The set of physical references and metadata instances used for visualisation
constitute the types of the network, each type being seen either as
a reference or a facet of the information space. This allows - as
it will be explained in the next section - the retrieval of co-occurrences
in one facet, based on one reference type - which can differ from
the physical reference.

\subsection{Multisets and hb-graphs}

Co-occurences can be seen as collaborations and therefore constitute
with their links a network. A collaboration is a $m$-adic relationship
as mentioned in \cite{newman2001scientific} between occurrences,
and therefore modelisation is often done with hypergraphs, i.e. family
of sets over a vertex set. But hypergraphs don't support neither hyperedge-based
repetition nor hyperedge-based weighting of vertices. We introduce
hb-graphs in \cite{ouvrard2018adjacency,ouvrard2019exthbgraphdiffusion,Ouvrard2019HbGraphMod}
to extend the concept of hypergraphs to families of multisets with
same universe, called the vertex set.

Multisets - also known as bags or msets - have been used for a long
time in many domains, in particular in text representation \cite{ouvrard2019exthbgraphdiffusion}.
A \textbf{multiset $A_{m}$} on a \textbf{universe} $A$ is a couple
$\left(A,m\right)$ where $A$ is a set and $m$ is an application
from $A$ to $\mathbb{W}\subseteq\mathbb{R}$ called the \textbf{multiplicity
function} of the multiset $A_{m}$. Elements that have a nonzero multiplicity
are gathered in the support of the multiset, written $A_{m}^{\star}.$
If the range of the multiplicity function is a subset of $\mathbb{N}$,
the multiset is called a \textbf{natural multiset}. In this case,
elements of the multiset can be seen as a non-ordered list of repeated
elements. The empty mset of universe $A$, written $\emptyset_{A}$,
is the multiset of empty support on the universe $A$.

Several notations of msets exist. Among the common notations mentioned
in \cite{ouvrard2018adjacency}, we note in this article a mset $A_{m}$
of universe $A=\left\{ x_{i}:i\in\left\llbracket n\right\rrbracket \right\} $
by: 
\[
A_{m}=\left\{ x_{i}^{m_{i}}:i\in\left\llbracket n\right\rrbracket \right\} 
\]
 where $m_{i}=m\left(x_{i}\right)$\footnote{A natural multiset can also be expressed as an unordered list with
repetition:
\[
A_{m}=\left\{ \left\{ \underset{m_{1}\,\text{times}}{\underbrace{x_{1},\ldots,x_{1}}},\ldots,\underset{m_{n}\,\text{times}}{\underbrace{x_{n},\ldots,x_{n}}}\right\} \right\} .
\]
}.

Different operations are defined between two multisets $A_{m_{A}}$
and $B_{m_{B}}$ of same universe $U$. Especially, as it is used
in this paper, the additive union of $A_{m_{A}}$ and $B_{m_{B}}$
is the multiset $C_{m_{C}}=A_{m_{A}}\uplus B_{m_{B}}$ of universe
$U$ such that for all $x\in U$, $m_{C}\left(x\right)=m_{A}\left(x\right)+m_{B}\left(x\right).$

Considering a set $V=\left\{ v_{1},...,v_{n}\right\} $ of vertices,
a \textbf{hb-graph }$\mathcal{H}=\left(V,E\right)$ is a family of
multisets $E=\left(e_{i}\right)_{i\in\left\llbracket p\right\rrbracket }$having
same universe $V$ and with support a subset of $V$. The elements
of $E$ are called the \textbf{hb-edges}. As a multiset, each hb-edge
$e_{i}\in E$ has a multiplicity function associated to it: $m_{e_{i}}:V\rightarrow\mathbb{W}$
where $\mathbb{W}\subset\mathbb{R}^{+}$. For a general hb-graph,
each hb-edge has to be seen as a weighted system of vertices, where
the weights of each vertex are hb-edge dependent.

When the multiplicity range of each hb-edge is a subset of $\mathbb{N}$
the hb-graph is said \textbf{natural}. A \textbf{hypergraph} is a
natural hb-graph where every vertex in any hb-edge has a binary value
- 0 or 1 - for multiplicity.

Considering the family of hb-edge support of a hb-graph $\mathcal{H}=\left(V,E\right)$,
we can define its \textbf{support hypergraph} as the hypergraph $\underline{\mathcal{H}}=\left(V,\underline{E}\right)$,
where $\underline{E}=\left\{ e^{\star}:e\in E\right\} $. The support
hypergraph is unique for a given hb-graph. But reconstructing the
hb-graph from a support hypergraph generates an infinite number of
hb-graphs, showing that the information contained in a hb-graph is
denser than in a hypergraph.

We represent hb-graphs using an unnormalised extra-node representation
\cite{ouvrard2018adjacency}: an extra-node per hb-edge is added and
the link thickness between it and each hb-edge support vertex is proportional
to the vertex multiplicity in this hb-edge.  The hypergraph support
of the hb-graph constitues a simplified representation.

\section{Hb-graph framework}

\label{sec:Hb-graph framework}

Multidimensional datasets are formed of data that are linked with
physical entities. For instance, a publication, a person, a piece
of music are possible physical entities. Some metadata of various
kind and types, numerical or not, are attached to the physical entities,
including their own reference.

Statistics can be easily performed on numerical types. For non numerical
types, only co-occurence gathering is easily achievable with traditional
charts and arrays. But choosing one of these non-numerical types as
a reference to build co-occurences enhances navigation in the information
space. The navigation is a simplification of the one presented in
\cite{Ouvrard2018ENDMHypergraphMod}. The hb-graph framework extends
the facet visualisation achieved in \cite{Ouvrard2018ENDMHypergraphMod}
to support multisets instead of sets using interconnected hb-graphs
at the level of the data instances called visualisation hb-graphs.

\subsection{Enhancing navigation}

Traditional database structures can be seen as hypergraphs where the
hyperedges reflect the table headers and the vertices the metadata
instances. Normalized forms of such databases are linked to properties
of the schema hypergraph \cite{fagin1983degrees}. In graph databases,
the schema\footnote{although not required \cite{McColl:2014}} represents
the relationships between the vertex types. The \textbf{schema hypergraph
}$\mathcal{H}_{\text{Sch}}=\left(V_{\text{Sch}},E_{\text{Sch}}\right)$
represents these relationships as hyperedges.

If database knowledge extraction processing is performed, such as
natural language processing, the schema hypergraph becomes an \textbf{extended
schema hypergraph} $\overline{\mathcal{H}_{\text{Sch}}}=\left(\overline{V_{\text{Sch}}},\overline{E_{\text{Sch}}}\right)$.

Some of the types in the extended schema have no interest neither
for visualisation, nor for being used as reference: we consider the
types of interest as a subset $U$ of $\overline{V_{\text{Sch}}}$
from which we generate the \textbf{extracted extended schema hypergraph}
$\mathcal{H}_{X}=\left(V_{X},E_{X}\right)$ where $V_{X}=U$, $E_{X}=\left\{ e\cap U:e\in\overline{E_{\text{Sch}}}\right\} .$

Only vertices of $\mathcal{H}_{X}$ that are \textbf{reachable} -
i.e. vertices with a simple path in between them - can be further
navigated and used either as reference or visualisation type. Therefore,
we build the \textbf{reachability hypergraph }$\mathcal{H}_{R}=\left(V_{R},E_{R}\right)$
with $V_{R}=V_{X}$ as its vertex set, individual hyperedges of $\mathcal{H}_{R}$
being the connected components $E_{\text{cc}}\left(\subset V_{X}\right)$
of $\mathcal{H}_{X}$. The hyperedges of $\mathcal{H}_{R}$ are not
connected as the connected components of a hypergraph constitute a
partition of its vertex set. When the reachability hypergraph has
only one hyperedge, the whole dataset is navigable: it is the ideal
situation.

We make the assumption that in each of the hyperedge of the reachability
hypergraph, it exists a metadata type or a combination of metadata
types that can be chosen as the \textbf{physical reference}. The data
instances related to this references are supposed to be unique. For
instance, in a publication dataset the physical reference is the id
of the publication itself.

Last hypergraph at the metadata level, the \textbf{navigation hypergraph}
$\mathcal{H}_{N}=\left(V_{N},E_{N}\right)$ is defined by choosing
a hyperedge $e_{r}\in E_{R}$ of the reachability hypergraph and a
non-empty subset $R_{\text{ref}}$ of $e_{r}$ of possible reference
types of interest. The choice of a subset $R$ of $R_{\text{ref}}$
allows to consider the remaining vertices of $e_{r}\backslash R$
as visualisation vertex types, that will be used to generate the facet
visualisation hb-graphs and are called the visualisation types. Hence:
$E_{N}=\left\{ e_{r}\backslash R:R\subseteq R_{\text{ref}}\land R\neq\emptyset\right\} .$
Navigation without changing references is possible only in one hyperedge
of $E_{N}$ at a time. The simplest case happens when there is only
one reference of interest selected at a time in $R_{\text{ref}}$;
we restrict ourselves to this case for the moment, i.e. we consider
for $E_{N}$ the set $E_{N/1}=\left\{ e_{r}\backslash R:R\subseteq R_{\text{ref}}\land\left|R\right|=1\right\} .$

In a publication dataset, typical metadata types are: \textit{publication
id}, title, abstract, \textit{authors}, affiliations, addresses, \textit{author
keyword}s, \textit{publication categories},\textit{ countries}, \textit{organisations},...\footnote{Metadata of interest for visualisation or referencing are in italic}
There are many different navigation hyperedge possible: for instance
choosing as reference the publication ids, the navigation hyperedge
is: \{authors, author keywords, organisations, country, publication
categories\}; choosing author keywords as reference the navigation
hyperedge is: \{authors, organisations, country, publication category,
publication ids\}.

\subsection{Facet visualisation hb-graphs}

Each physical entity $d$ in a dataset is described by a unique physical
reference $r$ and a set of data instances of different types $\alpha\in\overline{V_{\text{Sch}}}$.
In \cite{Ouvrard2018ENDMHypergraphMod}, we use sets to store co-occurrences.
Nonetheless in many cases, it is worth storing additional information
by joining a multiplicity - with nonnegative integer or real values
- to elements of co-occurences. For instance in a publication dataset,
different authors can have the same affiliation organization; retrieving
one occurence of organization per author enforces repetitions and
a natural multiset. When considering keywords, their relative frequency
in the document can be used as multiplicity. From \cite{ouvrard2019exthbgraphdiffusion},
we know that hb-graphs allow a refined ranking of the information.
Hence, the above facts motivate the usage of multisets to store co-occurrences.

We write $A_{\alpha,r}=\left\{ a_{1}^{m_{\alpha}\left(a_{1}\right)},...,a_{\alpha_{r}}^{m_{\alpha}\left(a_{\alpha_{r}}\right)}\right\} $
the multiset of values of type $\alpha$ - possibly empty - that are
attached to $d$, the physical entity. $d$ is entirely described
by its reference and the family of multisets that corresponds to co-occurrences
of the different types $\alpha$ in $\overline{V_{\text{Sch}}}$ linked
to the physical reference, i.e. $\left(r,\left(A_{\alpha,r}\right)_{\alpha\in\overline{V_{\text{Sch}}}}\right).$

Performing a search on the dataset retrieves a set $\mathcal{S}$
of physical references $r$. In the single-reference-restricted navigation
hypergraph, each hyperedge $e_{N}\in E_{N/1}$ describes accessible
facets relatively to a chosen reference type $\rho\in V_{N}\backslash e_{N}.$
Given a type $\alpha\in e_{N}$, the associated facet shows the visualisation
hb-graph $\mathcal{H}_{\alpha/\rho,\mathcal{S}}$ where the hb-edges
are the co-occurrences of type $\alpha$ relatively to reference instances
of type $\rho$ ($\alpha/\rho$ as short) retrieved from the different
references in $\mathcal{S}.$

We then build the co-occurrences $\alpha/\rho$ by considering the
set of all values of type $\rho$ attached to all the references $r\in\mathcal{S}$:
$\Sigma_{\rho}=\bigcup\limits _{r\in\mathcal{S}}A_{\rho,r}^{\star}.$
Each element $s$ of $\Sigma_{\rho}$ is mapped to a set of physical
references $R_{s}=\left\{ r:s\in A_{\rho,r}\right\} \in\mathcal{P}\left(\mathcal{S}\right)$
in which they appear: we write $r_{\rho}$ the mapping. The multiset
of values $e_{\alpha,s}$ of type $\alpha$ relatively to the reference
instance $s$ is $e_{\alpha,s}=\biguplus\limits _{r\in R_{s}}A_{\alpha,r}.$

The \textbf{raw visualisation hb-graph} for the facet of type $\alpha/\rho$
attached to the search $\mathcal{S}$ is then defined as:
\[
\mathcal{H}_{\alpha/\rho,\mathcal{S}}\overset{\Delta}{=}\left(\bigcup\limits _{r\in\mathcal{S}}A_{\alpha,r}^{\star},\left(e_{\alpha,s}\right)_{s\in\Sigma_{\rho}}\right).
\]

Since some hb-edges can possibly point to the same sub-mset of vertices,
we build a reduced visualisation weighted hb-graph from the raw visualisation
hb-graph. To achieve it we define: $g_{\alpha}:s\mapsto e_{\alpha,s}$
and $\mathcal{R}$ the equivalence relation such that: $\forall s_{1}\in\Sigma_{\rho}$,
$\forall s_{2}\in\Sigma_{\rho}$: $s_{1}\mathcal{R}s_{2}\Leftrightarrow g_{\alpha}\left(s_{1}\right)=g_{\alpha}\left(s_{2}\right).$

Considering a quotient class $\overline{s}\in\Sigma_{\rho}\big/\mathcal{R}$\footnote{$\Sigma_{\rho}\big/\mathcal{R}$ is the quotient set of $\Sigma_{\rho}$
by $\mathcal{R}$}, we write $\overline{e_{\alpha,\overline{s}}}=g_{\alpha}\left(s_{0}\right)$
where $s_{0}\in\overline{s}$.

$\text{\ensuremath{\overline{E_{\alpha}}}}=\left\{ \overline{e_{\alpha,\overline{s}}}:\overline{s}\in\Sigma_{\rho}\big/\mathcal{R}\right\} $
is the support set of the multiset $\left\{ \left\{ e_{\alpha,s}:s\in\Sigma_{\rho}\right\} \right\} $:
$\overline{e_{\alpha,\overline{s}}}\in\overline{E_{\alpha}}$ is of
multiplicity $w_{\alpha}\left(\overline{e_{\alpha,\overline{s}}}\right)=\left|\overline{s}\right|$
in this multiset.

It yields: $\left\{ \left\{ e_{\alpha,s}:s\in\Sigma_{\rho}\right\} \right\} =\left\{ \overline{e_{\alpha,\overline{s}}}^{w_{\alpha}\left(\overline{e_{\alpha,\overline{s}}}\right)}:\overline{s}\in\mathcal{S_{\rho}}\big/\mathcal{R}\right\} $

Let $\tilde{g_{\alpha}}:\overline{s}\in\Sigma_{\rho}\big/\mathcal{R}\mapsto e\in\overline{E_{\alpha}}$,
then $\tilde{g_{\alpha}}$ is bijective. $\tilde{g_{\alpha}}^{-1}$
allows to retrieve the class associated to a given hb-edge; hence
the associated values of $\Sigma_{\rho}$ to this class - which will
be important for navigation. The references associated to $e\in\overline{E_{\alpha}}$
are $\bigcup\limits _{s\in\tilde{g_{\alpha}}^{-1}(e)}r_{\rho}\left(s\right).$
The \textbf{reduced visualisation weighted hb-graph} for the search
$\mathcal{S}$ is defined as $\mathcal{H}_{\alpha/\rho,w_{\alpha},\mathcal{S}}\overset{\Delta}{=}\left(\bigcup\limits _{r\in\mathcal{S}}A_{\alpha,r}^{\star},\overline{E_{\alpha}},w_{\alpha}\right).$

The hb-graph support hypergraphs can be used to retrieve the results
given for hypergraphs in \cite{Ouvrard2018ENDMHypergraphMod}.

\subsection{Navigability through facets}

\begin{figure}
\begin{center}
\fontsize{12pt}{12pt}\selectfont
\begin{tikzpicture}[->,>=stealth',scale=0.9, font=\footnotesize, every node/.append style={transform shape}]

%***************
% physical entity
%***************
\node[state=blue,
		minimum width=3cm,
		minimum height=1cm] (phi) {
			\begin{tabular}{c}
				Physical entity: $d$\\
				Reference: $r$
			\end{tabular}};

%***************
% search set
%***************
\node[state=red!50!black,
		right of=phi,
		yshift=-1.5cm,
		minimum width=4cm,
		minimum height = 1.2cm,
		node distance=0cm] (search) {
		};

\node[] (search_Title) at ([xshift=-1.5cm,yshift=-0.7cm]search.north) {
			\begin{tabular}{c}
				$\mathcal{S}$\\
			\end{tabular}
		};
\node[state=blue,minimum width=1cm] (data1) at ([xshift=1em]search_Title.east) {\begin{tabular}{c}
				$d_1$\\
				$r_1$
			\end{tabular}};
\node[] (dataetc) at ([xshift=1em]data1.east) {$\ldots$};
\node[state=blue,minimum width=1cm] (datar) at ([xshift=1em]dataetc.east) {\begin{tabular}{c}
				$d_\mathcal{S}$\\
				$r_\mathcal{S}$
			\end{tabular}
		};

%***************
% alpha set
%***************
\node[state=green!50!black,
		right of=phi,
		yshift=0cm,
		minimum width=1.5cm,
		minimum height = 3.2cm,
		node distance=-4cm] (Aalpha) {
		};

\node[] (Aalpha_Title) at ([yshift=-0.5em]Aalpha.north) {
			\begin{tabular}{c}
				$A_{\alpha,r}$\\
			\end{tabular}
		};
\node[state=orange,minimum width=1cm] (aalpha1) at ([yshift=-1em]Aalpha_Title.south) {$a_{\alpha_1}$};
\node[] (aetc) at ([yshift=-0.5em]aalpha1.south) {\vdots};
\node[state=orange,minimum width=1cm] (aalphar) at ([yshift=-1em]aetc.south) {$a_{\alpha_r}$};

%***************
% rho set
%***************
\node[state=red!50!white,
		right of=phi,
		yshift=0cm,
		minimum width=1.5cm,
		minimum height = 3.2cm,
		node distance=4cm] (Arho) {
		};

\node[] (Arho_Title) at ([yshift=-0.5em]Arho.north) {
			\begin{tabular}{c}
				$A_{\rho,r}$\\
			\end{tabular}
		};
\node[state=orange,minimum width=1cm] (arho1) at ([yshift=-1em]Arho_Title.south) {$a_{\rho_1}$};
\node[] (arhoetc) at ([yshift=-0.5em]arho1.south) {\vdots};
\node[state=orange,minimum width=1cm] (arhor) at ([yshift=-1em]arhoetc.south) {$a_{\rho_r}$};

%***************
% alpha_prime set
%***************
\node[state=green!50!black,
		right of=phi,
		yshift=1.5cm,
		minimum width=4cm,
		minimum height = 1cm,
		node distance=0cm] (Aalphap) {
		};

\node[] (Aalphap_Title) at ([xshift=-1.5cm,yshift=-0.5cm]Aalphap.north) {
			\begin{tabular}{c}
				$A_{\alpha^\prime,r}$\\
			\end{tabular}
		};
\node[state=orange,minimum width=1cm] (aalphap1) at ([xshift=1em]Aalphap_Title.east) {$a_{\alpha^\prime_1}$};
\node[] (aalphapetc) at ([xshift=1em]aalphap1.east) {$\ldots$};
\node[state=orange,minimum width=1cm] (aalphapr) at ([xshift=1em]aalphapetc.east) {$a_{\alpha^\prime_r}$};

%**********************
% Result
%**************************

\draw[blue,line width=0.25mm,dotted] [<-]  (phi.south) to[*|] (data1.north);
\draw[blue,line width=0.25mm,dotted] [<-] (phi.south) to[*|] (datar.north);
\draw[green!50!black,line width=0.25mm] [-] (phi) -- (aalpha1);
\draw[green!50!black,line width=0.25mm] [-] (phi) -- (aalphar);
\draw[green!50!black,line width=0.25mm] [-] (phi) -- (arho1);
\draw[green!50!black,line width=0.25mm] [-] (phi) -- (arhor);
\draw[green!50!black,line width=0.25mm] [-] (phi) -- (aalphap1);
\draw[green!50!black,line width=0.25mm] [-] (phi) -- (aalphapr);
\end{tikzpicture}
\end{center}

\caption{Navigating between facets of the information space}

\label{Fig: Navigating_facets}
\end{figure}
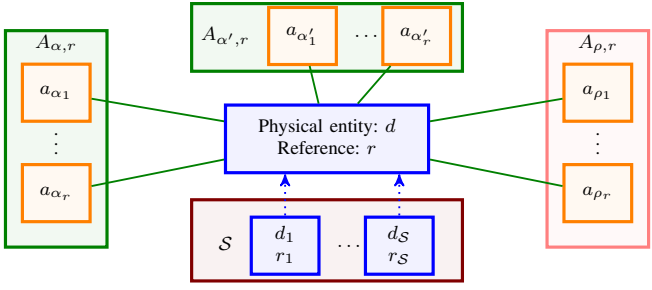

As for a given search $\mathcal{S}$ and a given reference $\rho$,
the sets $\Sigma_{\rho}$ and $R_{s},s\in\Sigma_{\rho}$ are fixed,
the navigability can be ensured between the different facets. We consider
a type $\alpha$, its visualisation hb-graph $\mathcal{H}_{\alpha/\rho,w_{\alpha}}$
and a subset $A$ of the vertex set of $\mathcal{H}_{\alpha/\rho,w_{\alpha}}$.
We target another type $\alpha'$ of co-occurrences referring to $\rho$
to be visualised. We illustrate the navigation in Figure \ref{Fig: Navigating_facets}.

{\footnotesize{}test}{\footnotesize\par}

We suppose that the user selects elements of $A$ as vertices of interest
from which s$\cdot$he wants to switch facet. Hb-edges of $\overline{E_{\alpha}}$
which contains at least one element of $A$ are gathered in $\left.\overline{E_{\alpha}}\right|_{A}=\left\{ e:e\in\overline{E_{\alpha}}\land\left(\exists x\in e:x\in A\right)\right\} .$
Using the application $\tilde{g_{\alpha}}^{-1}$ we retrieve the corresponding
class of references of type $\rho$ associated to the elements of
$\left.\overline{E_{\alpha}}\right|_{A}$, to build the set of references
$\left.\overline{V}\right|_{A}$ of type $\rho$ involved in the building
of co-occurences of type $\alpha'.$ Each of the classes in $\left.\overline{V}\right|_{A}$
contains instances of type $\rho$ that are gathered in a set $\mathcal{V}_{\rho,A}.$
Each element of $\mathcal{V}_{\rho,A}$ is linked to a set of physical
references by $r_{\rho}.$ Hence we obtain the physical reference
set involving elements of $A$: $\mathcal{S}_{A}=\bigcup\limits _{s\in\nu_{\rho,A}}R_{s}.$

The raw visualisation hb-graph $\left.\mathcal{H}_{\alpha'/\rho}\right|_{A}=\left(\bigcup\limits _{r\in\mathcal{S}_{A}}A_{\alpha',r}^{\star},\left(e_{\alpha',s}\right)_{s\in\mathcal{V}_{\rho,A}}\right)$
in the targeted facet is now enhanced using $\mathcal{S}_{A}$ as
search set $\mathcal{S}.$ To obtain the reduced weighted version
we use the same approach as above. The multiset of co-occurrences
retrieved includes all occurrences that have co-occurred with the
references attached to one of the elements of $A$ selected in the
first facet. Of course if $A=A_{\alpha,S}$ the reduced visualisation
hb-graph contains all the instances of type $\alpha'$ attached to
physical entities of the search $\mathcal{S}$.

The reference type can always be shown in one of the facet as a visualisation
hb-graph, that is in fact an hypergraph where all the hb-edges are
constituted of the reference itself in multiplicity the number of
time the reference occurs in the hb-graph.

\begin{figure}
\begin{center}

\begin{tabular}{>{\centering}m{3.8cm}>{\centering}m{3.8cm}}
\includegraphics[scale=0.15]{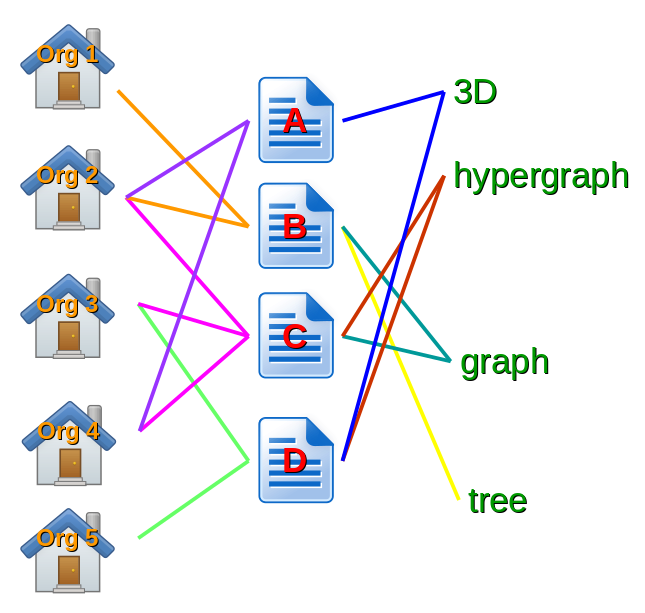} & \includegraphics[scale=0.15]{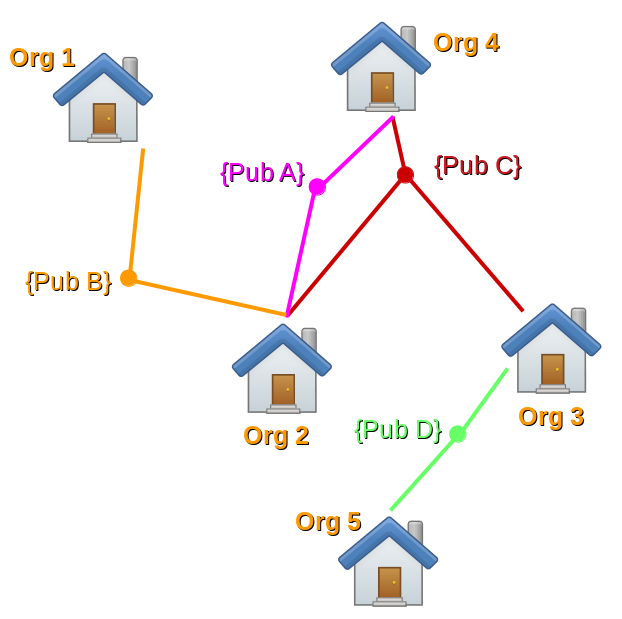}\tabularnewline
{\footnotesize{}(a)} & {\footnotesize{}(b)}\tabularnewline
\includegraphics[scale=0.15]{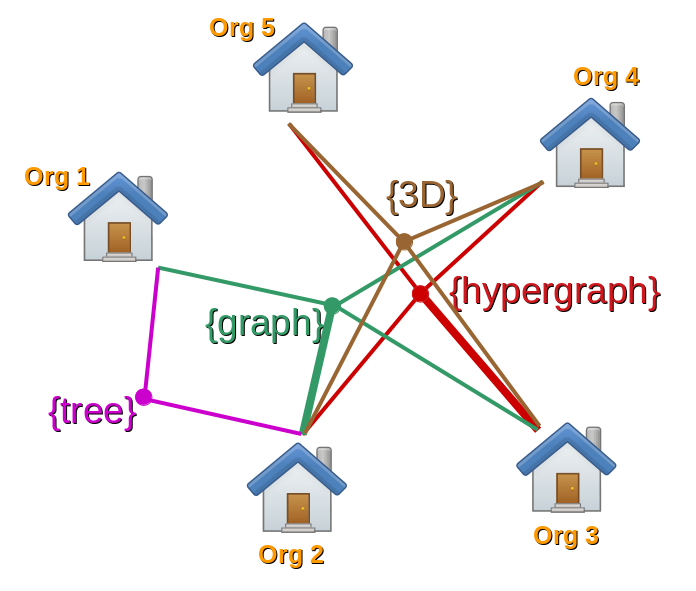} & \includegraphics[scale=0.15]{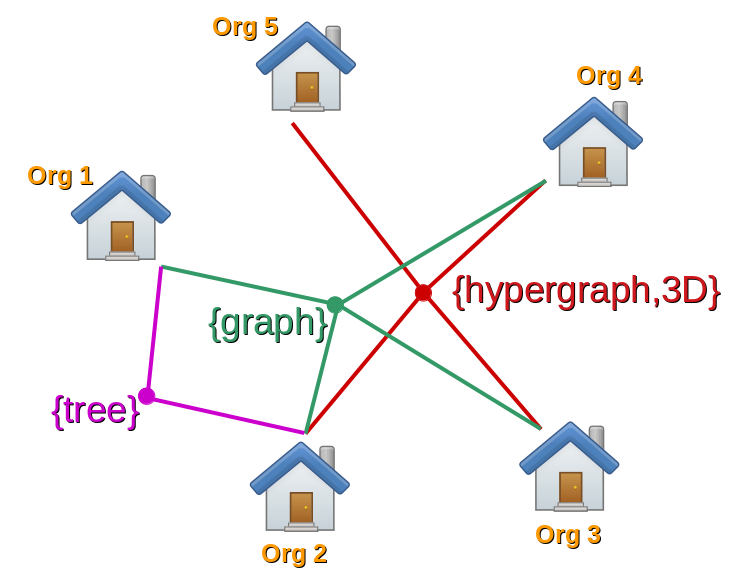}\tabularnewline
{\footnotesize{}(c)} & {\footnotesize{}(d)}\tabularnewline
\end{tabular}

\end{center}

{\footnotesize{}(a)~A publication network.}{\footnotesize\par}

{\footnotesize{}(b)~Reference: publication; Facet: organization;}{\footnotesize\par}

{\footnotesize{}View: hb-graph extra-node representation.}{\footnotesize\par}

{\footnotesize{}(c) and (d)~Reference: keywords; Facet: organization;}{\footnotesize\par}

{\footnotesize{}~~~~~(c)~View: hb-graph extra-node representation.}{\footnotesize\par}

{\footnotesize{}~~~~~(d)~View: hb-graph support hypergraph extra-node
representation.}{\footnotesize\par}

\caption{A co-occurence network and some visualisation hb-graphs}

\label{Fig:Visualisation hypergraph}
\end{figure}

Ultimately, by building a multi-dimensional network organized around
types, one can retrieve very valuable information from combined data
sources. This process can be extended to any number of data sources
as long as they share one or more types. Otherwise the reachability
hypergraph is not connected and only separated navigations are possible.
Figure \ref{Fig:Visualisation hypergraph} shows some examples of
visualisation hypergraphs.

\subsection{The DataHbEdron}

The DataHbEdron provides soft navigation between the different facets
of the information space. It has been introduced in \cite{Ouvrard2018ENDMHypergraphMod}
for hypergraphs and its principle is similar with the hb-graph support.

\begin{figure}
\begin{center}\includegraphics[scale=0.75]{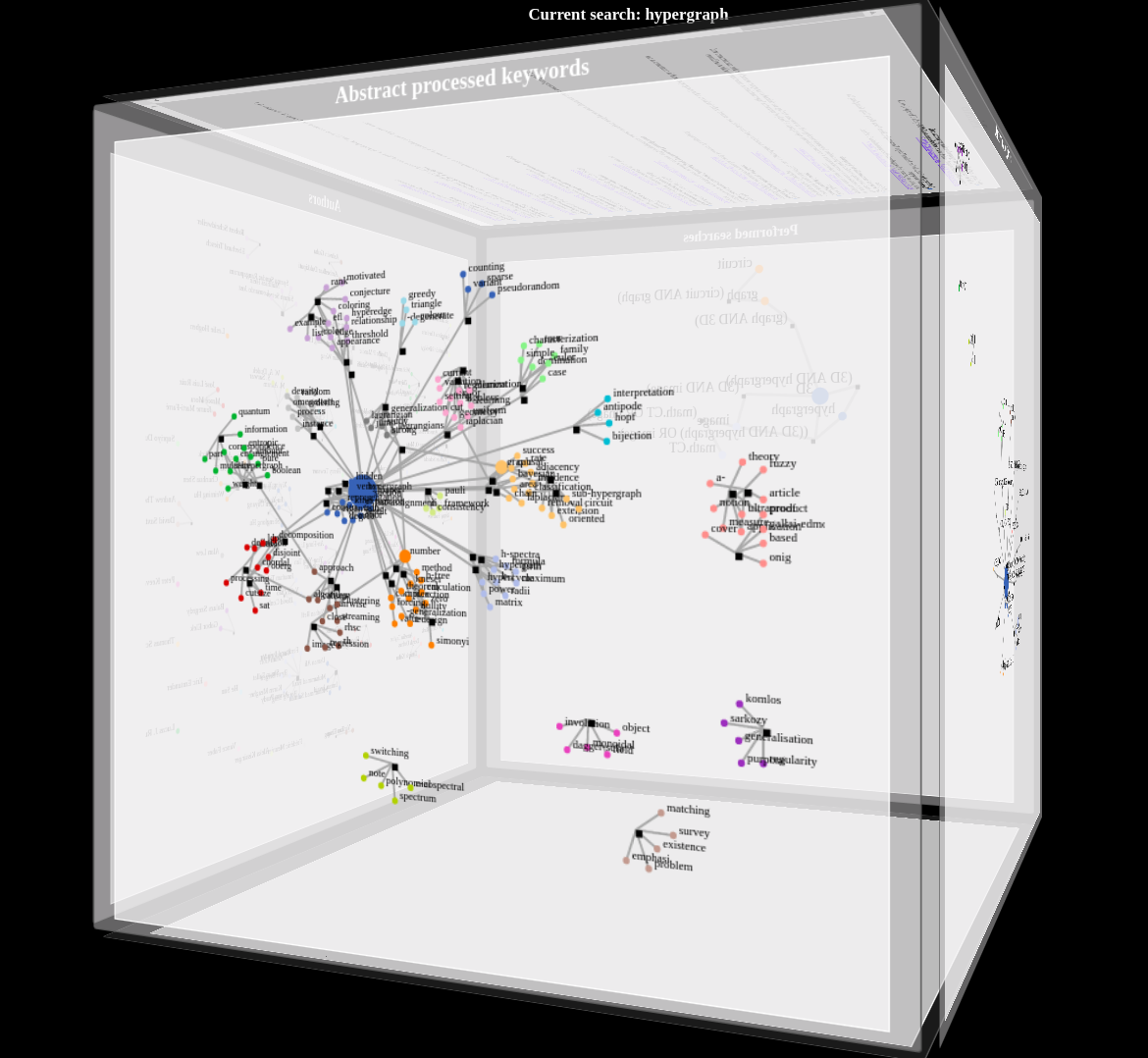}

\end{center}

\caption{DataHbEdron: cube shape}
\label{Fig: DataHbEdron}
\end{figure}

Each facet of the information space corresponding to a visualisation
type includes a visualisation hb-graph viewed in its 2D extra-node
representation with a normalised thickness on hb-edges \cite{ouvrard2019MCCCC}.
The different facets are embedded in a 2.5D representation called
the DataHbEdron. The DataHbEdron can be toggled between a cube - Figure
\ref{Fig: DataHbEdron} - and a carousel shape to ease the navigation
between facets. The reference facet is presented as a list of references
corresponding to the search output.

In the DataHbEdron, the faces show different facets of the information
space: the underlying visualisation hb-graphs allow as previously
explained the navigability through facets. Hb-edges are selectable
interactively between the different facets; as each hb-edge is linked
to a subset of the references, the corresponding references can be
used to highlight information in the different facets as well as in
the face containing the reference visualisation hb-graph.

\section{Results}

\label{sec:Results}

We applied this framework to perform searches and visual queries on
the Arxiv database. The results are visualised in the DataHbEdron
allowing simultaneous visualisation of the different facets of the
information space constitutes of authors, extracted keywords and subject
categories. The tool developped is now part of the Collaboration Spotting
family\footnote{\href{http://collspotting.web.cern.ch/}{http://collspotting.web.cern.ch/}}.
When performing a search, the standard Arxiv API\footnote{\href{https://arxiv.org/help/api/index}{https://arxiv.org/help/api/index}}
is used to query the Arxiv database. The queries can be formulated
either by a text entry or done interactively directly using the visualisation:
queries include single words or multiple words, with AND, OR and NOT
possible operators and parenthesis groupings. The querying history
is stored and presented as an interactive hb-graph to allow the visual
construction of complex queries including refinement of the queries
already performed. Each time a new query is formulated, the corresponding
metadata is retrieved by the Arxiv API.

When performing a search on Arxiv, the query is transformed into a
vector of words. The most relevant documents are retrieved based on
a similarity measure between the query vector and the word vectors
associated to individual documents. Arxiv relies on Lucene's built-in
Vector Space Model of information retrieval and the boolean model\footnote{\href{https://lucene.apache.org/core/2_9_4/scoring.html}{https://lucene.apache.org/core/2\_9\_4/scoring.html}}.
The Arxiv API returns the metadata associated to the document with
highest scores for the query performed. We keep only the first $n$
answers, with $n$ tunable by the end user. This metadata, filled
by authors during their submission of a preprint, contains different
information such as authors, Arxiv categories and abstract.

\begin{figure}
\begin{center}

\includegraphics[scale=0.15]{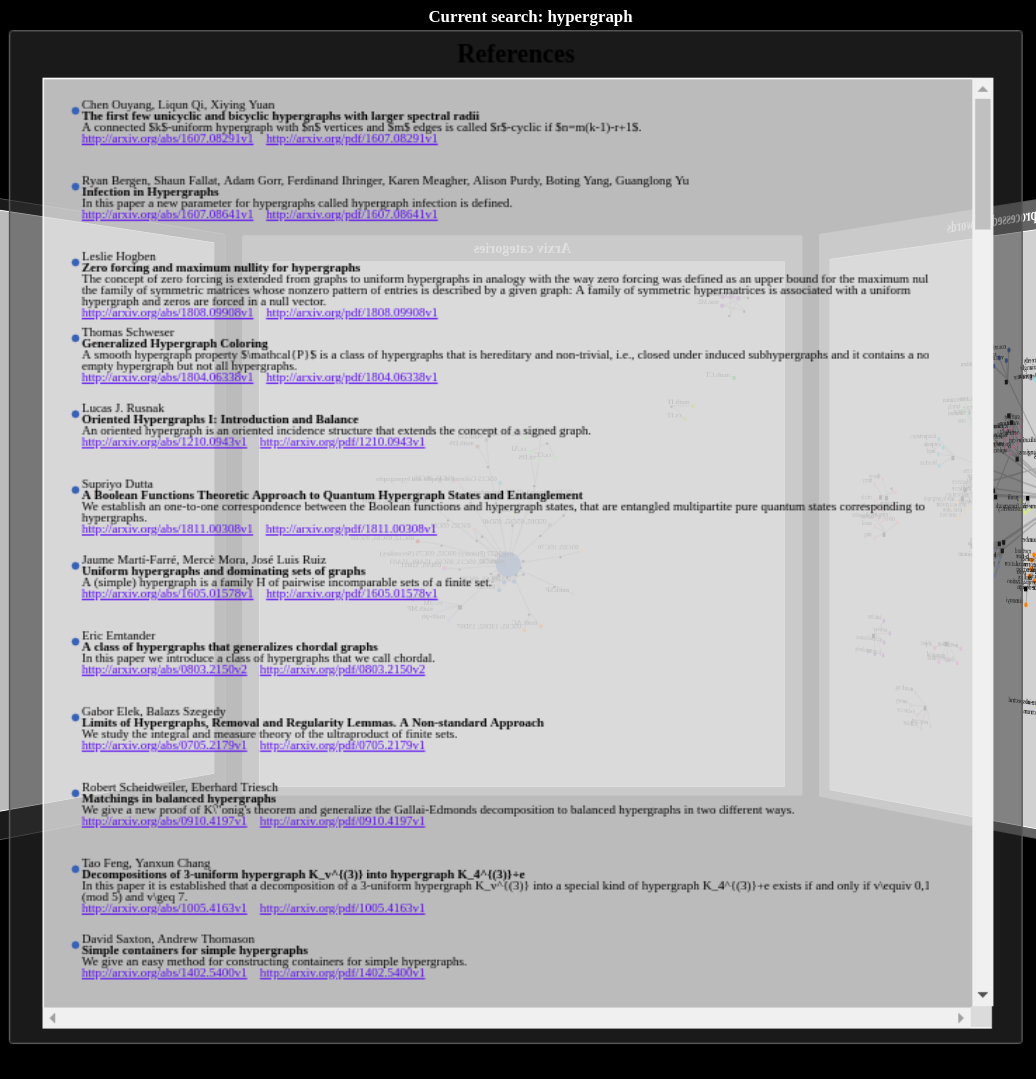}

\end{center}

\caption{First facet of the DataEdron: a well-known like classical verbatim
interface}
\label{Fig:DataEdron verbatim interface}
\end{figure}

The information space contains four main facets: the first facet shows
the Arxiv reference visualisation hb-graph with a contextual sentence
related to the query, links to Arxiv article's presentation and pdf.
This first facet layout is similar to classical textual search engines
- Figure \ref{Fig:DataEdron verbatim interface}.

The second facet corresponds to co-authors of the articles using as
reference the publication. The third facet depicts the co-keywords
extracted from the abstracts. The fourth facet shows the Arxiv categories
involved in the references.

Co-keywords are extracted from the abstracts using TextBlob, a natural
language processing Python library\footnote{\href{https://textblob.readthedocs.io/en/dev/}{https://textblob.readthedocs.io/en/dev/}}.
We extract only nouns using the tagged text, which has been lemmatized
and singularized.

Nouns in the abstract of each document are scored with TF-IDF, the
Term Frequency - Invert Document Frequency. Scoring each noun in each
abstract of the retrieved documents generates a hb-graphs $\mathcal{H}_{Q}$
of universe the nouns contained in the abstracts. Each hb-edge contains
a set of nouns extracted from a given abstract with a multiplicity
function that represents the TF-IDF score of each noun. In order to
limit the hb-edge size, we keep only the first $w$ words related
to an abstract, where $w$ is tunable by the end-user.

The fifth facet shows the queries that have been performed during
the session: the graph of those queries can be saved. The sixth facet
is reserved to show additional information such as the pdf of publications.

Any node on any facet is interactive, allowing to highlight information
from one facet to another by showing the hb-edges that are mapped
through the references. Queries can be build using the vertices of
the hb-graph, either isolated or in combination with the current search
using AND, OR and NOT through keyboard shortcuts and mouse. The first
query is the only one to be performed by typing it. Merging queries
between different users is immediate as they correspond to hb-edges
of a hb-graph. Queries are evolutive, gathered, stored and resketchable
months later.

\begin{figure}
\begin{center}

\includegraphics[scale=0.3]{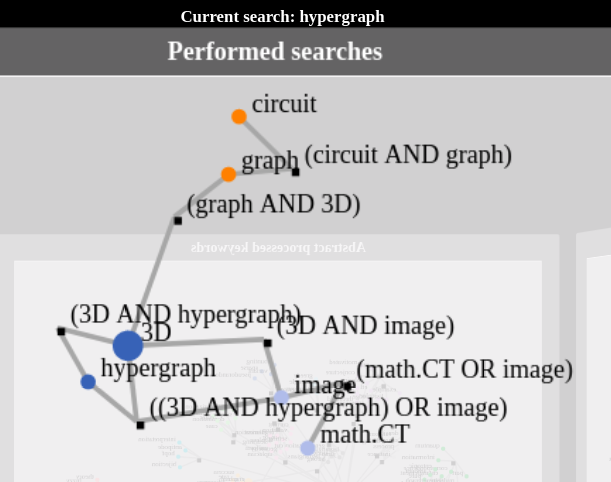}

\end{center}

\caption{Performed search}
\label{Fig:DataEdron search graph}
\end{figure}

The surfer has the possibility to display additional information related
to authors using dblp, to keywords using DuckDuckGo for disambiguation
and Wikipedia.

\section{Future work and Conclusion}

The framework presented in this paper supports visual query of datasets:
it enables full navigability of the corresponding of the corresponding
information space. It provides powerful insights into datasets using
simultaneous facet visualisation of the information space constructed
from the query results. The 2.5D visualisation helps to understand
the links in the dataset. This framework is flexible enough to enhance
user insight into many other multimedia content. Nonetheless, the
evaluation of such a human computer interface, as well as the evaluation
of the hb-graph visualisation remain open questions for experts of
these fields.

\label{sec:FW}

\section*{Acknowledgments}

This work is part of the PhD of Xavier OUVRARD, done at UniGe, co-supervised
by Pr. Stéphane MARCHAND-MAILLET and Dr Jean-Marie LE GOFF, head of
the Collaboration Spotting Project. The research is founded by a doctoral
position at CERN. The authors want to thanks Tullio BASAGLIA from
the CERN Library for his very precious advices and feedbacks on the
interface.

\bibliographystyle{ieeetr}
\bibliography{/home/xo/cernbox/these/000-thesis_corpus/biblio/references}

\begin{thebibliography}{10}

\bibitem{ranganathan1962elements}
S.~R. Ranganathan, {\em Elements of library classification}.
\newblock 1962.

\bibitem{ouvrard2019exthbgraphdiffusion}
X.~Ouvrard, J.-M. {Le Goff}, and S.~Marchand-Maillet, ``Diffusion by exchanges
  in hb-graphs: Highlighting complex relationships extended version,'' {\em
  Under submission}, 2019.

\bibitem{Ouvrard2018ENDMHypergraphMod}
X.~Ouvrard, J.~{Le Goff}, and S.~Marchand-Maillet, ``Hypergraph modeling and
  visualisation of complex co-occurence networks,'' {\em Electronic Notes in
  Discrete Mathematics}, vol.~70, pp.~65--70, 2018.
\newblock TCDM 2018 -- 2nd IMA Conference on Theoretical and Computational
  Discrete Mathematics, University of Derby.

\bibitem{dork12PivotPaths:StrollingthroughFacetedInformationSpaces}
M.~D{\"o}rk, N.~H. Riche, G.~Ramos, and S.~Dumais, ``Pivotpaths: Strolling
  through faceted information spaces,'' {\em IEEE Transactions on Visualization
  and Computer Graphics}, vol.~18, no.~12, pp.~2709--2718, 2012.

\bibitem{zhao2013interactive}
J.~Zhao, C.~Collins, F.~Chevalier, and R.~Balakrishnan, ``Interactive
  exploration of implicit and explicit relations in faceted datasets,'' {\em
  IEEE Transactions on Visualization and Computer Graphics}, vol.~19, no.~12,
  pp.~2080--2089, 2013.

\bibitem{agocs2017interactive}
A.~{Agocs}, D.~{Dardanis}, J.-M. {Le Goff}, and D.~{Proios}, ``Interactive
  graph query language for multidimensional data in collaboration spotting
  visual analytics framework,'' {\em ArXiv e-prints}, Dec. 2017.

\bibitem{han2011data}
J.~Han, J.~Pei, and M.~Kamber, {\em Data mining: concepts and techniques}.
\newblock Elsevier, 2011.

\bibitem{friedman1997bias}
J.~H. Friedman, ``On bias, variance, 0/1---loss, and the
  curse-of-dimensionality,'' {\em Data mining and knowledge discovery}, vol.~1,
  no.~1, pp.~55--77, 1997.

\bibitem{weber1998quantitative}
R.~Weber, H.-J. Schek, and S.~Blott, ``A quantitative analysis and performance
  study for similarity-search methods in high-dimensional spaces,'' in {\em
  VLDB}, vol.~98, pp.~194--205, 1998.

\bibitem{indyk1998approximate}
P.~Indyk and R.~Motwani, ``Approximate nearest neighbors: towards removing the
  curse of dimensionality,'' in {\em Proceedings of the thirtieth annual ACM
  symposium on Theory of computing}, pp.~604--613, ACM, 1998.

\bibitem{aggarwal2001surprising}
C.~C. Aggarwal, A.~Hinneburg, and D.~A. Keim, ``On the surprising behavior of
  distance metrics in high dimensional space,'' in {\em International
  conference on database theory}, pp.~420--434, Springer, 2001.

\bibitem{newman2001scientific}
M.~E. Newman, ``Scientific collaboration networks. ii. shortest paths, weighted
  networks, and centrality,'' {\em Physical review E}, vol.~64, no.~1,
  p.~016132, 2001.

\bibitem{ouvrard2018adjacency}
X.~Ouvrard, J.-M. {Le Goff}, and S.~Marchand-Maillet, ``Adjacency and tensor
  representation in general hypergraphs. part 2: Multisets, hb-graphs and
  related e-adjacency tensors,'' {\em arXiv preprint arXiv:1805.11952}, 2018.

\bibitem{Ouvrard2019HbGraphMod}
X.~Ouvrard, J.~{Le Goff}, and S.~Marchand-Maillet, ``Hb-graph modeling and
  visualisation of complex co-occurence networks,'' {\em Article under
  writing}, 2019.

\bibitem{fagin1983degrees}
R.~Fagin, ``Degrees of acyclicity for hypergraphs and relational database
  schemes,'' {\em Journal of the ACM}, vol.~30, no.~3, pp.~514--550, 1983.

\bibitem{McColl:2014}
R.~C. McColl, D.~Ediger, J.~Poovey, D.~Campbell, and D.~A. Bader, ``A
  performance evaluation of open source graph databases,'' PPAA '14,
  pp.~11--18, ACM, 2014.

\bibitem{ouvrard2019MCCCC}
X.~Ouvrard, J.-M. {Le Goff}, and S.~Marchand-Maillet, ``On hb-graphs and their
  application to general hypergraph e-adjacency tensor,'' {\em Article under
  submission to the MCCCC32 proceedings}, 2019.

\end{thebibliography}

\end{document}